\documentclass[preprint,authoryear,12pt]{elsarticle}

\usepackage{graphics}

\journal{Annals of Nuclear Energy}

\begin{document}

\begin{frontmatter}



\title{Steady-State Neutronic Analysis of Converting the\\UK CONSORT Reactor for ADS Experiments}


\author[man1]{Hywel Owen} \corref{cor1}\ead{hywel.owen@manchester.ac.uk}
\author[man2]{Matthew Gill}
\author[ic]{Trevor Chambers}
\cortext[cor1]{Corresponding author. Tel: +44 1925 603797}

\address[man1]{Cockcroft Accelerator Group, School of Physics and Astronomy, University of Manchester, Manchester M13 9PL, UK}
\address[man2]{Nuclear Physics Group, School of Physics and Astronomy, University of Manchester, Manchester M13 9PL, UK}
\address[ic]{Imperial College London, London SW7 2AZ, UK}

\begin{abstract}
CONSORT is the UK's last remaining civilian research reactor, and its present core is soon to be removed. This study examines the feasibility of re-using the reactor facility for accelerator-driven systems research by replacing the fuel and installing a spallation neutron target driven by an external proton accelerator. MCNP5/MCNPX were used to model alternative, high-density fuels and their coupling to the neutrons generated by 230~MeV protons from a cyclotron striking a solid tungsten spallation target side-on to the core. Low-enriched $\textrm{U}_\textrm{3}\textrm{Si}_{\textrm{2}}$ and U-9Mo were considered as candidates, with only U-9Mo found to be feasible in the compact core; fuel element size and arrangement were kept the same as the original core layout to minimise thermal hydraulic and other changes. Reactor thermal power up to 2.5~kW is predicted for a $k_{eff}$ of 0.995, large enough to carry out reactor kinetic experiments.

\end{abstract}


\end{frontmatter}



\section{ADS Experiments}

Accelerator-driven systems (ADS) are subcritical reactors in which the external neutron source may allow operation with inherent safety (\cite{Degweker:2007dp}), and the improved neutron economy enables applications in fuel breeding (particularly for Thorium-based fuels) and for actinide management to reduce long-lived waste (\cite{Wilson:wn}, \cite{Bowman:1992in}, \cite{Daniel:1996hf}, \cite{Nifenecker:1999el}). The so-called 'Energy Amplifier' has been proposed as a combination of all these features in a subcritical lead-cooled fast reactor with solid fuel elements, in which fast neutrons are generated in a single central spallation target and delivered to a core-blanket arrangement within which actinide burning is achieved via slow crossing of absorption resonances during small-lethargy scatters (\cite{Carminati:1993uh}, \cite{Rubbia:1995ux}).

Whilst system ADS designs such as MYRRHA at SCK-CEN are well-progressed (\cite{Abderrahim:2001ws}), and several key components such as the central lead spallation target have been proven in scaled-down prototypes (\cite{Groeschel:2004wi}), there is as yet little practical experience with accelerator-reactor coupling experiments. The first true accelerator-coupled experiment has only recently been carried out at KURRI with a very low current (1~nA) of protons into the KUCA A core (\cite{Shiroya:2000vy}, \cite{Shiroya:2001th}, \cite{Shiroya:2002vo}, \cite{Tanigaki:2004vl}), and there have been several ADS studies incorporating D-D or D-T neutron generators, particularly the MUSE (\cite{Soule:2004wb}, \cite{Billebaud:2007vk}), GUINEVERE (\cite{Mercatali:2010kc}, \cite{Uyttenhove:2011in}) and YALINA experiments (\cite{Persson:2005uk}). Also, there are the well-developed TRADE studies of using the Rome TRIGA reactor (\cite{Naberejnev:2003vx}). However, there has been as yet no demonstration of ADS operation at intermediate powers between these initial experiments at only a few W(th) power, and the 100-3000 MW(th) powers envisaged for full ADS operation. In this paper we consider how to produce an intermediate power output of up to 3~kW(th) in an existing light water research reactor, using a commercial cyclotron as a proton driver for a spallation target. This would enable studies of reactor kinetics relevant to high power fast systems (\cite{Naberejnev:2003vx}, \cite{HerreraMartinez:2004tk}, \cite{HerreraMartinez:2007ky}) and the development of neutron diagnostic methods required to safely operate at these powers. 

Conceptual ADS designs often consider reactor powers of up to 1~GW(e) output, in which a 1~GeV proton beam drives spallation efficiently in a lead target, with $\sim$20 neutrons per proton (n/p) being a typical production ratio (\cite{Lone:1995tv}). Safety considerations mean that subcritical $k_{eff}$ values between 0.95 and 0.98 are typically considered (\cite{Nifenecker:1999el}), which combined with the reactor output implies beam powers in excess of 5~MW. Whilst designs for proton linacs exist that deliver such powers (in particular the designs considered for the European Spallation Source, see for example \cite{Lengeler:1998ip} and \cite{Lindroos:2011hh}), to date there has been no demonstration of the reliability required for ADS operation, which is limited both by worries about the robustness of the target and core under rapid thermal cycling (see for example \cite{Takei:2009vf}), and about the economic feasibility of power interruptions from a nuclear power plant based on ADS (\cite{Steer:2011jn}).

The realisation that reliable, high power accelerators are a limiting factor in constructing a viable ADS power plant has resulted in three responses. The first response has been the significant research into replacement technologies for high power linacs, in particular the use of Fixed-Field, Alternating Gradient (FFAG) accelerators as a way of overcoming the energy limitation of cyclotrons (\cite{Tanigaki:2004vl}). The recent success of the proof-of-principle EMMA accelerator (\cite{Barlow:1352706}) indicates that this non-scaling variant of the FFAG could potentially enable high-power (10~MW) low-cost protons at the required 1~GeV energy, but we should also note that researchers involved with the DAEDALUS neutrino project (\cite{Alonso:2010yz}, \cite{Conrad:2010fo}) are examining novel cyclotron designs (\cite{Calanna:2011vx}) as an alternative method to meet the twin requirements of power and reliability, and which may also be applied to ADS (\cite{Kim:2001vj}). Even with lower-cost cyclotron or FFAG designs, space charge effects in the accelerated proton bunches limit the beam current and the required high reliability is considered difficult to achieve (\cite{Craddock:2008tm}). Multiply-redundant designs have therefore been considered in which three accelerators deliver protons to independent targets within the ADS core (\cite{Takizuka:1999ta}, \cite{Broeders:2000vs}), the idea being that a failure in one accelerator can be made up for by increasing the power of the other two.

The second response is to consider operation at higher $k_{eff}$ values, perhaps as high as $k_{eff}=0.998$, which alleviates the beam power requirements but of course raises questions over whether the reactor can remain safe kinetically, or as fuel burnup proceeds. Aker/Jacobs have proposed the 'ADTR' as a potential design (\cite{Fuller:2010vz}), but the safety will be determined by whether the $k_{eff}$ can be measured reliably at full power (using effectively a source-jerk method). However, even when operating close to $k_{eff}=1$ the envisaged power is still well over the c.~1~MW level where costs become high and reliability becomes difficult to achieve.

The third response is to lower the output power of the reactor from c.~1~GW(e) to values of 100~MW(e) or less. Whilst this alleviates the power requirement from a single accelerator, it does not solve the lack of reliability. But, if $k_{eff}$ can be increased in such a reactor to values closer to unity, there is the possibility to use multiple, 'off-the-shelf' accelerators - such as high power superconducting electron linacs driving photofission via a gamma-producing Bremsstrahlung target - that can deliver the required neutron flux in the core (\cite{Diamond:1999ij}). The use of multiple driver accelerators provides the redundancy required to achieve high reliability, but at a sufficiently low $k_{eff}$ the number of required accelerators and their overall wall-plug power may rise to a level where there is no net energy gain in the complete ADS system. Again, a very high $k_{eff}$ close to one will probably be required. With present technology it is also contested whether providing a particular electrical capacity with multiple small modular reactors (SMRs) can compete with the traditional large LWR plants, but there is considerable current interest in replacing the economies of scale of PWRs with the economies of mass-production that may be offered by SMRs (\cite{Anonymous:2007vl}).

The multiple-target accelerator-driven SMR (ADSMR) may hold some potential, and the use of multiple targets in addition may allow selective control of burnup without needing to shuffle fuel elements, perhaps also facilitating a 'sealed battery' operation which could help with proliferation resistance. But, as indicated above, there are a number of issues that should be addressed before such a scheme can be validated. The main questions are: can a high $k_{eff}$ ADS system be safely operated, using suitable monitoring of the subcriticality level; can multiple accelerators provide useful modification of the flux profile in a reactor core, over and above the efficiency advantages offered by burnable poisons. Fuel utilisation may be improved by the use of accelerators, but their additional cost should be considered.

To help answer the above questions, we have considered a modification of the existing CONSORT reactor to incorporate a spallation target driven by an external moderate-power ($\sim 1\mu \textrm{A}$ current) accelerator, which will enable an increase in the available external neutron flux at the core by several orders of magnitude compared to previous ADS experiments.

\section{The CONSORT Reactor}
\label{}

\begin{figure}
\begin{center}
    \includegraphics[width=80mm]{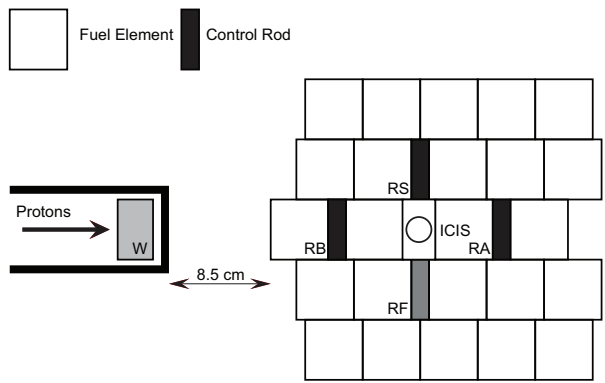}
    \caption{\label{corelayout}Schematic cross-section of the CONSORT reactor at the vertical core centre-line, showing the arrangement of the fuel elements with respect to the large irradiation tube that penetrates into the surrounding water volume. There are two other such irradiation tubes each above and below the central one, and the inner face of the tubes lies approximately 8.5~cm from the closest fuel element, meaning that significant neutron thermalisation occurs from the proposed 100~mm-diameter Tungsten spallation target in the irradiation tube. The much smaller $\sim25$~mm ICIS irradiation tube penetrates the centre of the core, but is too small for a target and associated beam transport over the 6~m distance to the outside of the reactor.}
\end{center}
\end{figure}

\begin{figure}
\begin{center}
    \includegraphics[width=70mm]{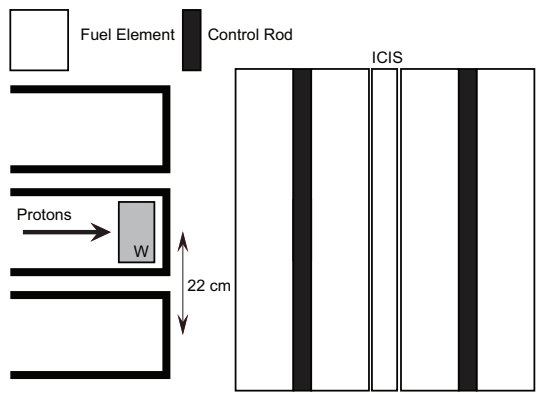}
    \caption{\label{irradiationtube}Schematic cross-section through the side-on irradiation tubes showing how the central tube may be initially used with a solid Tungsten spallation target.}
\end{center}
\end{figure}

CONSORT is a low-flux, 100 kW(th) civilian pool-type light-water research reactor that has been operated by Imperial College London since 1965 \citep{Grant:1965wf}; it is now the only remaining operating civilian research reactor in the UK, and there are plans to decommission the present core over the next few years. The present core consists of an optimised arrangement of 24 fuel elements (each approximately 3" square in cross-section) of roll-bonded U-Al MTR type. Three types of element are presently used: MARK I, II and III. In MARK I/II there are 12 curved plates per element, whilst in MARK III elements there are 16 flat plates; there is typically a 4 mm water gap between each plate. The reactor contains four control rods: 3 'coarse' Cadmium rods clad in stainless steel (one used as a safety rod); and one 'fine' Stainless Steel rod. The complete core assembly is approximately 400~x~400~x~600~mm, is located in an Aluminium reactor vessel approximately 6~m deep, and is surrounded by a concrete enclosure with graphite reflectors with several penetrations to enable insertion of samples for neutron irradiation (see Figure \ref{corelayout}). Of particular interest for potential ADS studies are the central $\sim25$~mm ICIS irradiation tube, and the three larger side-on tubes that penetrate into the reactor vessel to lie quite close to one side of the core (see Figure \ref{irradiationtube}).

We have considered the re-use of as much of the present facility as possible (buildings, shielding and vessel) to allow an ADS experiment at modest cost. We examined replacement fuel with the same disposition as the present core arrangement so that thermal management in the core would be expected to be similar; whilst the control rod mechanisms would have to be replaced, we envisage them to retain their original arrangement.

Although the central ICIS irradiation tube can deliver protons to the centre of the core (similar to the scheme proposed for TRADE in \cite{Naberejnev:2003vx}), we consider that the narrow $\sim25$~mm diameter and long length 6~m throw to the outside of the reactor vessel makes beam transport difficult to a target. Also, the small reactor vessel diameter means there is insufficient space, both between the control mechanisms and above the reactor (1.94 m to the bottom of the fuel-handling crane), to insert a beam transport system; the associated shielding at high level in the reactor building is not considered to be practicable. Instead we propose the use of the three side-on irradiation tubes which penetrate close to the core face, which allow for a much larger c.~100~mm-diameter target to be used. We envisage a beam transport/target plug that may be conveniently inserted initially into the central tube, the upstream beam transport connecting to a proton source to be located in an adjacent, non-nuclear-licensed, building. This arrangement does not require any substantive changes to the existing building and reactor infrastructure, and as such should be simpler to license. The overall arrangement is similar to the side-on coupling adopted at KURRI/KUCA (\cite{Shahbunder:2010tr}), and we believe that the shielding arrangement from proton source to reactor can be fitted within the existing building.

\section{Fuel Choice and Core Modelling}

\begin{figure}
\begin{center}
    \includegraphics[width=65mm]{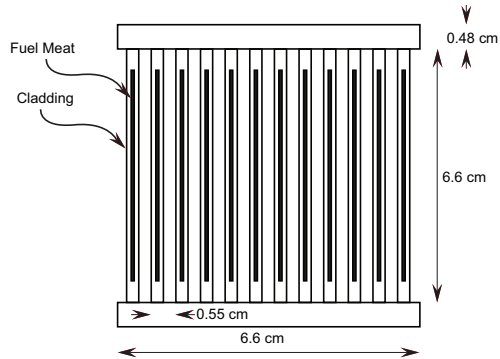}
    \caption{\label{fuelelement}Fuel assembly cross-section as used in the MCNP/MCNPX modelling. The fuel meat dimensions vary from plate to plate, but we consider average values of 0.5~mm meat thickness and 1.5~mm overall plate thickness to be adequate.}
\end{center}
\end{figure}

To enable use of the current core geometry, we considered $\textrm{U}_\textrm{3}\textrm{Si}_{\textrm{2}}$ and U-9Mo as candidate fuels, using a similar fuel meat/Al cladding to the present core and adopting the same plate arrangement and separation (see Figure \ref{fuelelement}). Modelling of the core was performed using MCNP5 (\cite{Anonymous:2003vi}), and the spallation and core-target coupling using MCNPX (v2.6.0) (\cite{Anonymous:2008ul}); both codes were used with ENDF/B-VII.0 cross-section data (\cite{Chadwick:2006bm}), and spallation reactions were modelled using the Bertini intra-nuclear cascade method (see discussion below), which we believe is sufficiently accurate for the present study. The core and control rods, water moderator, Aluminium vessel, Graphite reflector, irradiation tubes and simplified support mechanisms were included in the model, which was found in previous studies to give $k_{eff}$ estimates sufficiently close to measured values (\cite{Jiang:2006tw}).

\begin{figure}
\begin{center}
    \includegraphics[width=80mm]{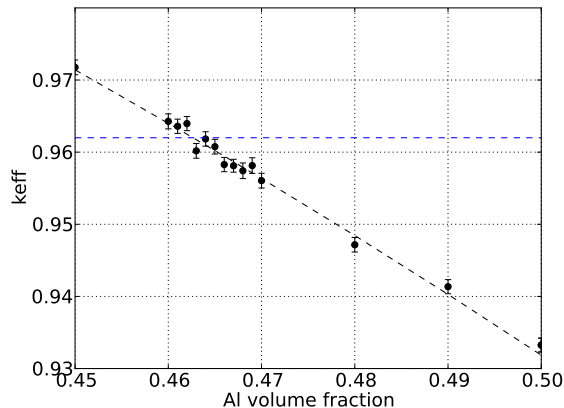}
    \caption{\label{usimix}Variation in $k_{eff}$ with fuel meat Aluminium fraction for $\textrm{U}_\textrm{3}\textrm{Si}_{\textrm{2}}$ fuel with 19.9\% enrichment, when all core control rods are inserted. The present core $k_{eff}$ under the same conditions is shown as a blue dotted line.}
\end{center}
\end{figure}

It is thought that the lowest feasible volume of Aluminium in the fuel meat is 55\% when $\textrm{U}_\textrm{3}\textrm{Si}_{\textrm{2}}$ is used as a fuel dispersant (\cite{Keiser:2003dz}). Even using 19.9\% enrichment, it is not possible to obtain sufficient reactivity in the core unless the Aluminium volume fraction is unfeasibly low (see Figure \ref{usimix}). $\textrm{U}_\textrm{3}\textrm{Si}_{\textrm{2}}$ is therefore unsuitable as a fuel.

\begin{figure}
\begin{center}
    \includegraphics[width=90mm]{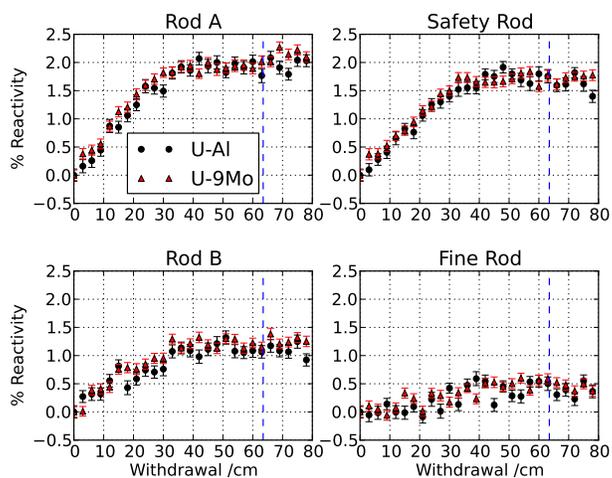}
    \caption{\label{rodworths}Effect on reactivity of withdrawing each control rod, comparing the present core (U-Al) with the proposed core (U-9Mo). The vertical blue dashed lines indicate full withdrawal of the control rods.}
\end{center}
\end{figure}

U-9Mo has a high Uranium density and has been shown to have good performance under irradiation even at high burnup (\cite{Snelgrove:1997bm}). Tests of U-9Mo (under the RERTR programme) have been carried out at Idaho National Laboratory using the same thickness plates as the ones we consider here (\cite{Wight:2008uz}). In the present CONSORT application the fuel temperature and burnup will remain low, so fuel-cladding interaction, swelling and failure are not thought to pose any significant problems, although must be considered in more detail (\cite{VandenBerghe:2010vx}). We varied Uranium enrichment and determined the required percentage of Aluminium in the fuel meat to give criticality with the control rods in their present operational position: a single coarse rod and the fine rod half-way in (30~cm), and the safety and other coarse rod fully withdrawn. There is a wide range of feasible enrichments when using U-9Mo; for later results we have chosen an Aluminium volume fraction of 75\%, corresponding to a fuel enrichment near the maximum of 19.5\%: a summary of the proposed new core properties is given in Table \ref{tab:coreproperties}. The variation of reactivity when withdrawing each control rod is shown in Figure \ref{rodworths}, and indicates that the core behaviour with U-9Mo fuel is sufficiently similar to the original U-Al fuel. The change in reactivity is also consistent with a different model of the core in MCNP developed separately \cite{Jiang:2006tw}. We therefore have confidence that a new core can, in principle, be operated similarly to the present core.

\begin{figure}
\begin{center}
    \includegraphics[width=80mm]{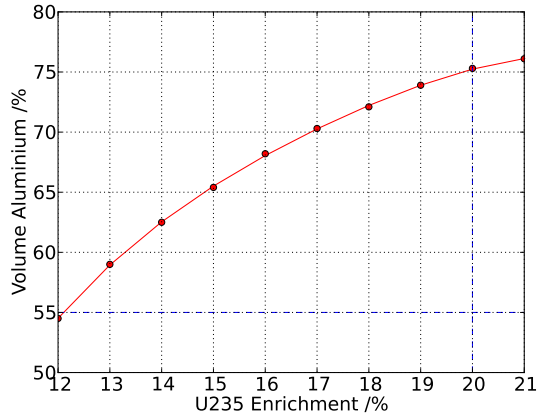}
    \caption{\label{u9moenrichment}Required Aluminium percentage in the fuel meat of a U-9Mo dispersed fuel element as a function of fuel enrichment. The horizontal blue dashed line shows the minimum possible Aluminium percentage (55\%), whilst the vertical blue dashed line shows the maximum permitted enrichment of 19.9\%.}
\end{center}
\end{figure}

\begin{table}
\caption{\label{tab:coreproperties}Reference core design with U-9Mo fuel used at 19.5\% enrichment.}
\begin{center}
\begin{tabular}{ll}
\hline
Fuel & U-9Mo/Al \\
Al-Fuel Mixing (by Volume) & $\sim$ 75 \% Al \\
Fuel Cladding & (Pure) Al \\
Number of Fuel Assemblies & 24 \\
Plates per Fuel Assembly & 12 \\
Uranium Enrichment & 19.5\% \\
Power & Max 100 kW(th) \\
Core Height & 0.63 m \\
Coolant/Moderator & Light Water \\
Reflector & Graphite \\
\hline
\end{tabular}
\end{center}
\end{table}

\section{Accelerator Proton Source}

A number of options are possible for generating neutrons in the core. Spallation becomes very efficient at higher proton energies above a few hundred MeV and, depending upon the (high-Z) target material and dimensions, a broad optimum exists around 1~GeV incident proton energy where there is an optimal trade-off between the energy expended to accelerate protons and the number of neutrons generated per proton. A 1~GeV proton beam generates typically over 20 neutrons per proton (n/p) from a lead target of sufficient length and radius (\cite{Lone:1995tv}, \cite{Hilscher:1998dh}). However, whilst this efficiency is required to generate the copious number of neutrons for a high-power, low $k_{eff}$ subcritical system, for a lower-power demonstrator it is much more cost-effective to reduce the proton energy. At very low energies nuclear reactions such as p on $^7$Li ((\cite{Kononov:2006ut}, Liskien:1975wx)) or p on $^9$Be  (\cite{Guzek:1998um}) may be used with high-current (several mA), low energy (several MeV) electrostatic or radio-frequency quadrupole (RFQ) proton sources to give neutron fluxes in excess of those delivered by D-T generators, and with better reliability and lifespan. Higher proton energies between 100-250~MeV - deliverable by commercial cyclotrons at currents up to 1~$\mu$A - allow spallation reactions to be used, albeit with rather poor efficiencies $\sim1$ n/p. In general terms, larger proton currents are available at lower energies, but this must be traded against the conversion efficiency to neutrons at those lower energies and the greater target power for a given output neutron flux.

For the CONSORT application we consider that a proton energy of 230-250~MeV with an associated current of 1~$\mu$A is sufficient. Whilst significantly lower power than the 110~MeV proton beam to be delivered for the TRADE experiment, it may be supplied by an already-demonstrated commercial cyclotron such as those supplied by IBA (230~MeV normal-conducting, see \cite{Jongen:2004vs}) or Varian (250~MeV superconducting, see \cite{Klein:2005tf} and \cite{Kim:2007un}). This also limits the target heating that must be dissipated in the confined (and therefore hard-to-cool) target plug region. 

It is worth noting that several alternatives to protons exist. Deuterons have been widely used, particularly in D-D RFQ neutron sources and electrostatic D-T generators. Whilst these generators deliver nearly-monoenergetic neutrons (which are beneficial in some applications) production rates are not superior to Li or Be targets, and lost deuterons and tritium use give rise to more difficult activation and radiological issues. However, for very low power initial experiments we would consider the use of a D-T generator (as has been used in initial experiments at KURRI and GENEPI) as such sources are at least an order of magnitude cheaper than higher-flux alternatives. At lower powers one may also consider the use of Bremsstrahlung-based targets, as mentioned earlier. In this case electrons are used to create broad-spectrum gamma rays peaked at an energy suitable for stimulating photofission, i.e. $E_\gamma\sim$~15~MeV (Wilke:1990gj), which requires electron energies of at least 30~MeV (Berger:1970fu). The advantage of such an approach in ADS systems is that that the gammas are more forward-directed than an equivalent neutron source. The efficiency production in a thick target (typically Tungsten or Tantalum) is perhaps as high as 1 gamma of suitable energy for every 3 electrons, but of course the photofission cross-sections for both $^{235}$U and $^{238}$U are rather low (peaking at approximately 400~mb and 160~mb respectively). Whilst very high electron currents up to 100~mA are possible at 50~MeV using superconducting cavities, the target power is also extremely high. Such sources are used to provide radioactive beams using Uranium target fission (\cite{Koscielniak:2008zz}), and have been proposed for the production of isotopes and for ADS systems (\cite{Liu:2005up}), but in our application where cooling is difficult, we consider that a medium-energy proton source of c.~230-250~MeV is better. The beam current of c.~1 $\mu$A commercially available at this energy results in a target power of 230-250~W which is manageable with air cooling, although there is space to provide a water circuit if required. We compare the overall expected neutron production yields with alternative candidate proton sources in Table \ref{targetyields}.

\begin{table*}
\caption{\label{targetyields}Neutron yields from differing targets using incident protons, scaled from existing data/simulations to deliver $1.125\times10^{13}$ n/s as desired for CONSORT. The current required for CONSORT is compared to that achieved (or proposed in the case of TRADE). Lower-energy protons result in an impractical target power, whilst the use of 1~GeV protons is discounted because of the size and capital cost of the accelerator, irrespective of the achievable current. A 140~MeV high current proton cyclotron is under development by ENEA/IBA, but is not yet demonstrated. We note that there is significant ($\times 4$) discrepancy in the expected $^7$Li(p,n)$^7$Be yield between \cite{Culbertson:2004vz} and \cite{Kononov:2006ut}. }
\begin{center}
\begin{tiny}
\begin{tabular}{llllllll}
\hline
Target & E /MeV & n/p  & I /$\mu$A (ach.) & I /$\mu$A (req.) &  Power /W & p/s & Reference \\
\hline
Li & 2.8 & 0.0009 & 1000 & 2000 & 5600 & $1.25\times 10^{16}$ & \cite{Culbertson:2004vz}* \\
Be & 30 & 0.019 & 60 to 250 & 96 & 2885 & $6.01\times 10^{14}$  & \cite{Abbas:2009ts} \\
W & 140 & 0.75 & (200-300) & 2.4 & 336 & $1.50\times 10^{13}$  & \cite{HerreraMartinez:2007ky} \\
W & 230 & 1.8 & 0.8 to 1.0 & 1 & 230 & $6.25\times 10^{12}$  & this work \\
Pb & 1000 & 20 & N/A & 0.09 & 90 & $5.625\times 10^{11}$  & \cite{Hilscher:1998dh} \\
\hline
\end{tabular}
\end{tiny}
\end{center}
\end{table*}

\section{Target Neutron Production and Core Coupling}

Following the selection of source/target combination we must determine the characteristics of neutron production in the target. Tungsten was chosen as a target material as it is robust, widely used, and well-understood. An air-cooled target at the moderate power level of 230-250~W should not experience significant damage, but there is the option to coat it with Tantalum if required, as is regularly done for high-power solid spallation targets (\cite{Broome:1996vs}, \cite{Nio:2005tr}, \cite{Findlay:2007ut}). Similarly, we have as yet only considered a preliminary bare unreflected target, as the water surrounding the irradiation tube acts to scatter some side-going neutrons toward the core. Later designs can in principle incorporate energy filtering and reflection, but we consider this a minor issue at present as the neutrons are in any event well-thermalised by the intervening water layer in this first scheme.

\begin{figure}
\begin{center}
    \includegraphics[width=80mm]{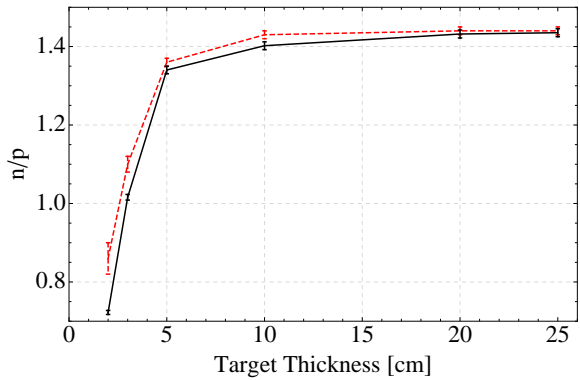}
    \caption{\label{pbtarget}Neutron multiplicity from 197~MeV protons in a Lead target of 12~cm diameter of varying thickness, comparing data taken from Lott et al. (dashed, red) with MCNPX simulations (solid, black).}
\end{center}
\end{figure}

\begin{figure}
\begin{center}
    \includegraphics[width=80mm]{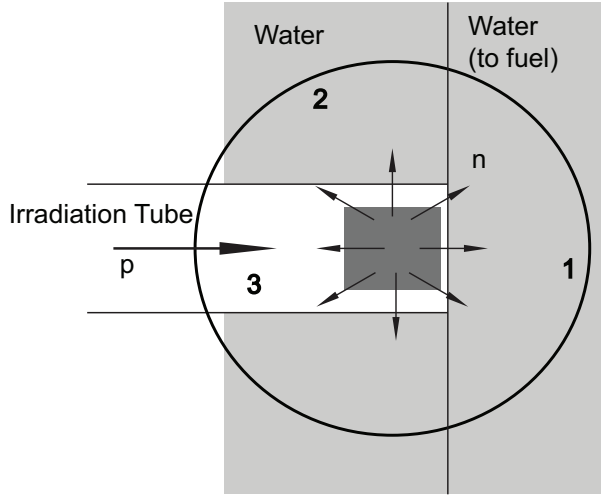}
    \caption{\label{regions}Definition of regions of forward-going (1), side-going (2) and backward-going (3) neutrons as used for the Tungsten target analysis.}
\end{center}
\end{figure}

\begin{figure}
\begin{center}
    \includegraphics[width=80mm]{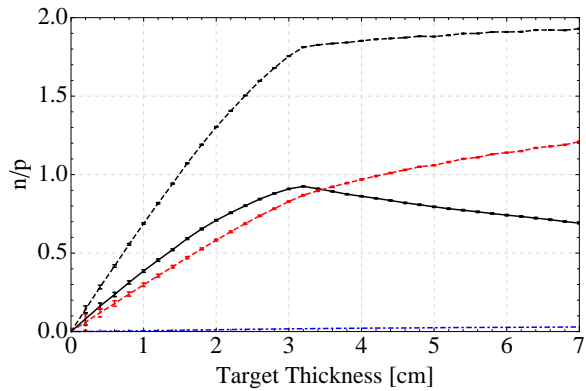}
    \caption{\label{relativeyields}Predicted neutron multiplicity in a Tungsten target with 230~MeV incident protons subdivided into forward-going (solid, black), side-going (dashed, red) and backward-going (blue, dashed) components as defined in Figure \ref{regions}, also showing the total neutron multiplicity. Error bars for the small backward-going component are suppressed for clarity. Self-absorption and scattering in the target results in a peak in the forward-going neutron production for thicknesses slightly less than that required to stop the protons.}
\end{center}
\end{figure}

We validated our MCNPX simulations near to the chosen proton energy by comparing the predicted neutron multiplicity against measured data in a lead target at 197~MeV (\cite{Lott:1998wl}); this is shown in Figure \ref{pbtarget}. The stopping distance for protons of 230/250~MeV is approximately 3.4/3.8~cm in Tungsten including straggling and was cross-checked in PSTAR (\cite{Berger:1992tt}) and SRIM2010/2011 (\cite{Ziegler:2010wt}). Whereas self-absorption in a Lead target is not significant, in Tungsten it can be: although longer targets give greater total neutron production, the increase is mostly from side-going neutrons which will be multiply-scattered in the water moderator and are less likely to reach the fuel. A tally sphere of radius 30~cm subdivided according to Figure \ref{regions} was used to determine the proportion of forward-going neutrons (past a surface co-planar with the target end), i.e. those neutrons most likely to contribute to fission in the core. The maximum forward-going component is achieved for target lengths slightly less than the stopping distance for the protons. We therefore chose a target thickness of 3.5~cm to ensure complete proton stopping in the target: roughly half the neutrons in this case are forward-going (see Figure \ref{relativeyields}). Note that in these simulations the proton beam at the target was assumed to be Gaussian with a diameter of 3~cm.

\begin{figure}
\begin{center}
    \includegraphics[width=90mm]{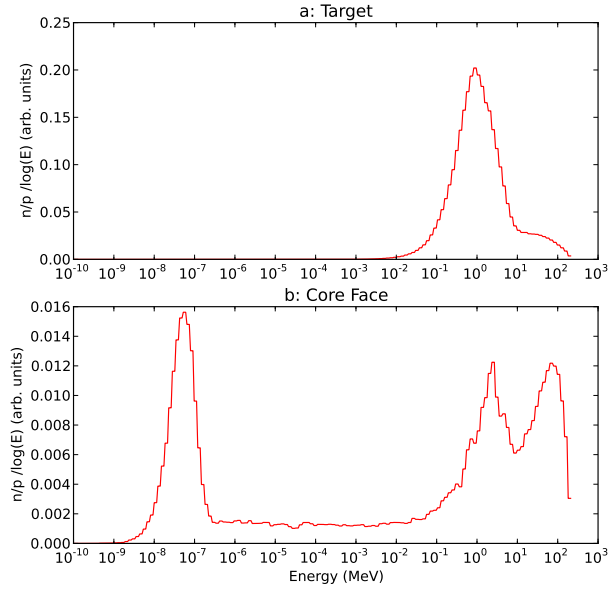}
    \caption{\label{spectrum}Thermalisation of neutrons from the spallation target to the core face due to the intervening water.}
\end{center}
\end{figure}

\begin{figure}
\begin{center}
    \includegraphics[width=80mm]{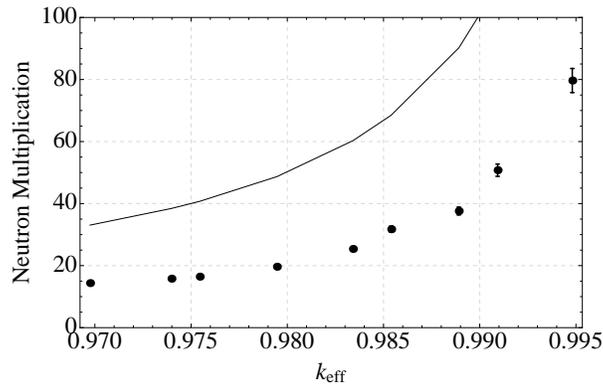}
    \caption{\label{combplot}Example neutron multiplication under subcritical operation for differing $k_{eff}$ values as all four control rods are withdrawn. The range of $k_{eff}$ shown is from all rods fully inserted to 16~cm withdrawn ($k_{eff}=0.995$). The multiplication is shown per target neutron reaching the fuel region (individual data points), and is compared to the expected multiplication for an equilibrium critical neutron distribution using KCODE (solid line).}
\end{center}
\end{figure}

\begin{figure}
\begin{center}
    \includegraphics[width=80mm]{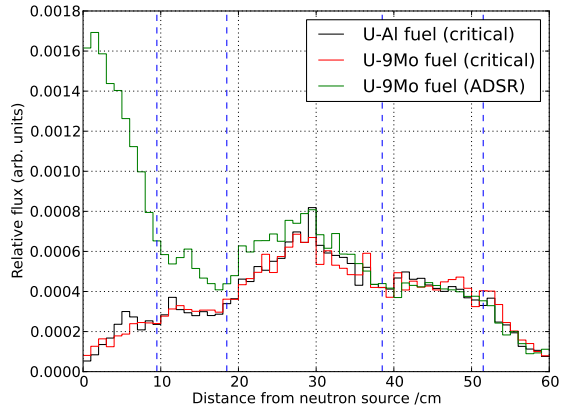}
    \caption{\label{fluxprofile}Lateral flux profile through the core under different operating scenarios. Comparing the original U-Al fuel with the proposed U-9Mo core we see no significant difference in flux profile, indicating that the core behaviour in critical operation will be similar. In subcritical operation there is a significant modification to the flux profile. The vertical dashed lines indicate proposed locations for neutron flux measurement using compact detectors.}
\end{center}
\end{figure}

MCNPX was used to calculate the coupling of the spallation target to the core: significant thermalisation occurs over the 8.5~cm between the target and the core face (Figure \ref{spectrum}), that distance being greater than the typical slowing-down length in water of 3~cm for the spallation neutrons (\cite{Nellis:1977tn}); approximately 0.18 neutrons per proton make it into the fuel, a little more than 10\% of those generated in the target. The resulting neutron multiplication of those neutrons is shown in Figure \ref{combplot}, and differs from that calculated using KCODE (\cite{Anonymous:2003vi}) since in the side-on target configuration the neutron distribution through the core is very asymmetric (Figure \ref{fluxprofile}).

\begin{figure}
\begin{center}
    \includegraphics[width=80mm]{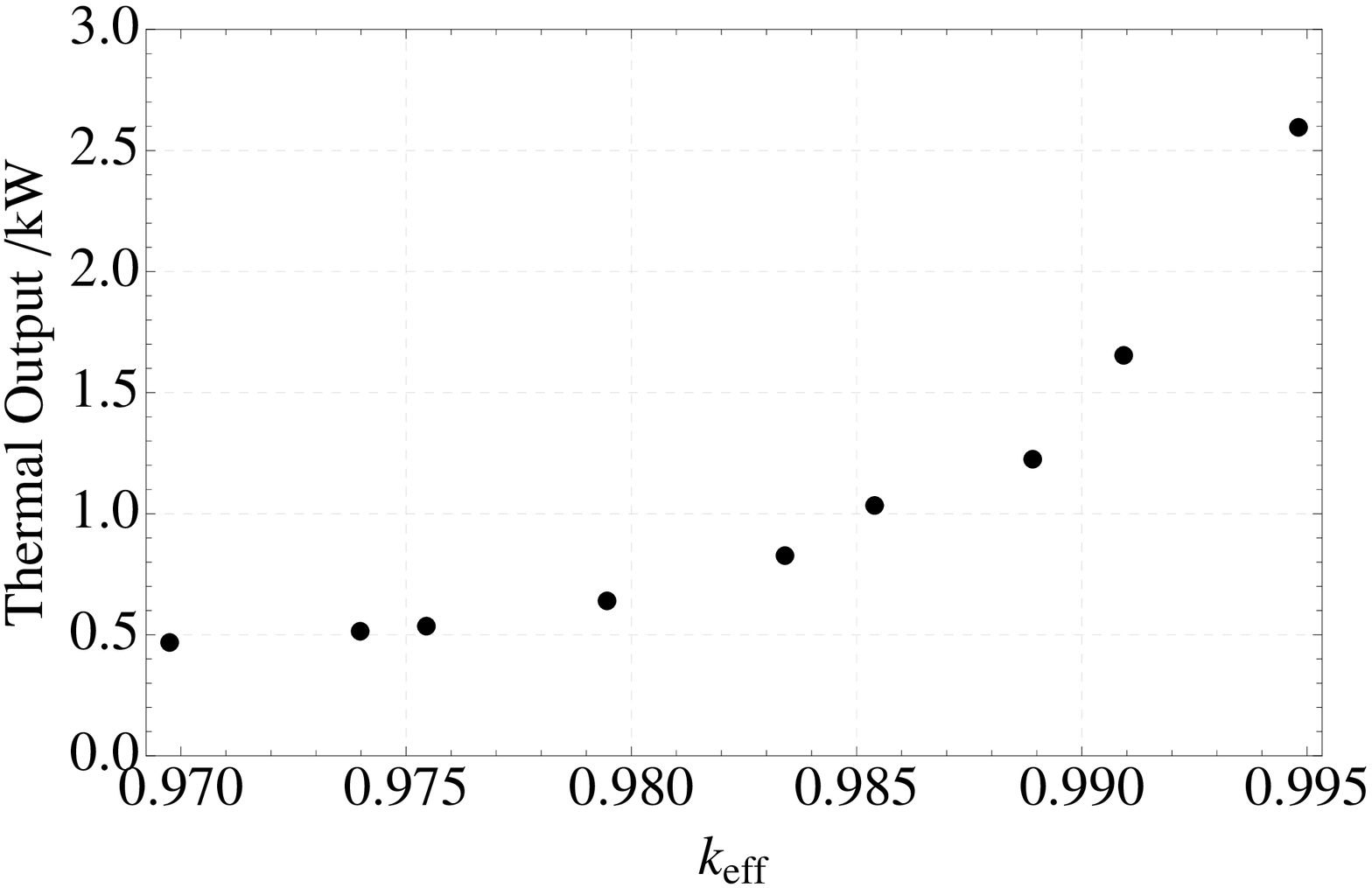}
    \caption{\label{powerplot}Reactor power variation with $k_{eff}$ for input 230~MeV proton beam current of $1\mu$A, assuming prompt energy release of 181~MeV per fission.}
\end{center}
\end{figure}

The predicted power generated in the core under ADS operation is given in Figure \ref{powerplot}. Thermal powers up to a few kW are possible even with a relatively low-efficiency coupling between the target and the core. Several improvements would in principle be possible, but have yet to be studied. Firstly (as mentioned above) reflectors could be placed around the target to improve the guiding of neutrons to the fuel; secondly, some voiding of the core region and intermediate water moderation could be made, analogous to the scheme adopted at KUCA for their ADS tests (\cite{Shahbunder:2010tr}). Together these could increase the neutron flux within the core and make the distribution more symmetric, increasing the available thermal power. However, the latter would require much more careful analysis of the thermal hydraulic issues.

\section{Discussion and Further Work}

The present steady-state analysis indicates that significant thermal power increases can be obtained if CONSORT were re-configured for ADS operation than has yet been demonstrated elsewhere; this enables medium-power feedback experiments to be carried out that will help determine whether operation closer to $k_{eff}=1$ is possible than previously considered. In common with other proposals we would also need a careful analysis of both the thermal hydraulic behaviour of the overall system under input changes and transients, and to accurately map the neutron flux through the core. We envisage the installation of small Silicon-based (\cite{DaVia:2008zza}, \cite{Caruso:2010ks}) or scintillator-based neutron monitors (\cite{Yamane:1999jt}, \cite{Yagi:2010ub}) to carry out flux measurements to characterise the differences in behaviour between critical and subcritical operation, and to measure the degree of subcriticality. These areas are subjects of ongoing study.

The overall power attainable in the experiment proposed here is less than that proposed for other experiments, for example at Dubna (\cite{Shvetsov:2006gd}, Gudowski:2006ht,), but is still several orders of magnitude greater than that achieved to date. The combination of re-use of existing reactor infrastructure and the use of a commercial cyclotron as proton driver reduces the complexity of delivering an experiment, and therefore should be significantly cheaper than other approaches.

\section{Acknowledgements}

We greatly appreciate the advice and expertise provided by David Bond (Imperial College Reactor Centre), and Matthew Eaton (Imperial College London, Department of Earth Sciences and Engineering). Matthew Gill was supported by a grant from the UK Nuclear FiRST Doctoral Training Centre, funded by Engineering and Physical Sciences Research Council.

\bibliographystyle{model2-names}








\end{document}